\documentclass[aps,pra,twocolumn,superscriptaddress,showpacs]{revtex4-1}

\usepackage{bm}
\usepackage{amsthm}
\usepackage{amsmath,amssymb}
\usepackage{graphicx}
\usepackage{braket}
\usepackage{enumerate}
\usepackage{ascmac}
\usepackage{mathrsfs}

\newcommand{\e}[1]{\begin{align} #1\end{align}}
\newcommand{\tr}[1]{\mathrm{tr}\left[ #1 \right]}
\newcommand{\h}[1]{\hat{#1}}
\newcommand{\pd}[3]{
 \if 1#1 \frac{\partial #2}{\partial #3}
 \else \frac{\partial^{#1} #2}{\partial #3^{#1}}\fi}
 \newcommand{\od}[3]{
 \if 1#1 \frac{{\mathrm d} #2}{{\mathrm d} #3}
 \else \frac{{\mathrm d}^{#1} #2}{{\mathrm d}#3^{#1}}\fi}
\newcommand{\calE}[0]{{\bm {\mathcal E}}}

\begin{document}


\title{Tradeoff Relation between Information and Disturbance in Quantum Measurement}


\author{Tomohiro Shitara}
\email[]{shitara@cat.phys.s.u-tokyo.ac.jp}
\affiliation{Department of Physics, University of Tokyo, 7-3-1 Hongo, Bunkyo-ku, Tokyo 113-8654, Japan}
\author{Yui Kuramochi}
\affiliation{Department of Nuclear Engineering,
Kyoto University, Kyoto 615-8540, Japan}
\author{Masahito Ueda}
\affiliation{Department of Physics, University of Tokyo, 7-3-1 Hongo, Bunkyo-ku, Tokyo 113-8654, Japan}
\affiliation{Center for Emergent Matter Science (CEMS), RIKEN, Wako, Saitama 351-0198, Japan}


\date{\today}

\begin{abstract}
When we extract information from a system by performing a quantum measurement, the state of the system is disturbed due to the backaction of the measurement.
Numerous studies have been performed to quantitatively formulate tradeoff relations between information and disturbance.
We formulate a tradeoff relation between information and disturbance from an estimation-theoretic point of view, and derive an inequality between them.
The information is defined as the classical Fisher information obtained by the measurement, and the disturbance is defined as the average loss of the quantum Fisher information. 
We show that pure and reversible measurements achieve the equality of the inequality.
We also identify the necessary condition for various divergences between two quantum states to satisfy a similar relation.
The obtained relation holds  not only for the quantum relative entropy but also for the maximum quantum relative entropy.
\end{abstract}

\pacs{}

\maketitle

\section{introduction}
When we perform a quantum measurement and extract information from a system, the system is disturbed due to the backaction of the measurement.
The more information we extract by quantum measurement, the more strongly the system is disturbed.
Such a tradeoff relation has been recognized since Heisenberg pointed out the uncertainty relation~\cite{He27} between error and disturbance.
Since 1990's, with the tremendous development of the quantum information theory, various methods of quantifying the information and the disturbance has been proposed, and the tradeoff relations has been shown based on their definitions\cite{Fu96,Da03,Sa06,BS06,BHH08,CL12}.
In most of the studies, the information and the disturbance are defined in the information theoretic settings, i.e., decoding the message encoded in quantum states by a measurement.

In this paper, we  formulate the information and the disturbance in the setting of estimating an unknown quantum state by quantum measurements, with an emphasis on the estimation accuracy.
We derive an inequality that shows the tradeoff relation between them, and give a sufficient condition to achieve the equality.
We also discuss the condition for divergences, which measure the distinguishability between two quantum states, to satisfy a similar inequality.

\section{information-disturbance relation based on estimation theory}
Suppose that we estimate an unknown quantum state corresponding to a density operator $\h\rho_{\bm\theta}$ by performing a quantum measurement.
Here, $\bm\theta\in\Theta\subset\mathbb R^m$ represents $m$ real parameters that characterize the unknown state, so that estimating the state is equivalent to estimating the parameters.
Such a parameterized family of states $\{\h\rho_{\bm\theta}\}_{\bm\theta\in\Theta}$ is called a quantum statistical model.
A quantum measurement is characterized by a mapping from quantum states to the probability distribution of outcomes and the post-measurement state corresponding to each outcome.
If the measurement outcome is discrete, the probability $p_{\bm\theta,i}$ of obtaining the outcome $i\in I$ and the post-measurement state $\h\rho_{\bm\theta,i}$ are respectively given by 
\e{
p_{\bm\theta,i}&=\sum_j \tr{\h K_{ij}\h\rho_{\bm\theta}\h K_{ij}^\dagger}\\
\h\rho_{\bm\theta,i}&=\frac{1}{p_{\bm\theta,i}}\sum_j \h K_{ij}\h\rho_{\bm\theta}\h K_{ij}^\dagger,
}
where measurement operators $\{\h K_{ij}\}$ satisfy the normalization condition $\sum_{i,j}\h K_{ij}^\dagger\h K_{ij}=\h I$.

What is the natural quantification of the information in the setting of estimating an unknown state from the outcome of the quantum measurement?
The estimating process is characterized by a function $\bm\theta^{\rm est}:I\rightarrow \Theta$ which is called an estimator.
Since the outcome $i\in I$ is a random variable, the estimator, which is calculated from the outcome, is also a random variable, and should be distributed around the true parameter $\bm\theta$.
According to the classical Cram\'er-Rao inequality, the variance-covariance matrix of the locally unbiased estimator has a lower bound which is determined only by a family of the probability distributions $\{p_{\bm\theta}\}_{\bm\theta\in\Theta}$:
\e{
{\rm Var}_{\bm\theta}[\bm\theta_{\rm est}]\ge (J^C_{\bm\theta})^{-1}.
}
Here, $J^C_{\bm\theta}$ is an $m\times m$ matrix called the classical Fisher information whose elements are defined as
\e{
[J^C_{\bm\theta}]_{ab}=\sum_i  p_{\bm\theta,i} \pd{1}{\log p_{\bm\theta,i}}{\theta_a} \pd{1}{\log p_{\bm\theta,i}}{\theta_b}. \label{info}
}
Therefore, a measurement with a larger classical Fisher information allows us to estimate the state more accurately. 
We define the classical Fisher information as the information obtained by the quantum measurement.

The disturbance can be evaluated as the loss of information on the parameter $\bm\theta$ that can be extracted from the quantum state.
The information on the parameter $\bm\theta$ that the quantum state potentially possesses can be quantified by the quantum Fisher information, which is defined as
\e{
[J^Q_{\bm\theta}]_{ab}:=\tr{\pd{1}{\h\rho_{\bm\theta}}{\theta_a}\bm K^{-1}_{\h\rho_{\bm\theta}} \pd{1}{\h\rho_{\bm\theta}}{\theta_b}}.
}
Here, the superoperator $\h K_{\h\rho}$ is defined as
\e{
\h K_{\h\rho}=\bm R_{\h\rho}f(\bm L_{\h\rho}\bm R_{\h\rho}^{-1}),
}
where $f:(0,\infty)\rightarrow (0,\infty)$ is an operator monotone function satisfying $f(1)=1$ and $\bm R_{\h\rho}  (\bm L_{\h\rho})$ is the right (left)-multiplication of $\h\rho$:
\e{
\bm R_{\h\rho}(\h A)=\h A\h\rho, \ \bm L_{\h\rho}(\h A)=\h\rho\h A.
}
The quantum Fisher information is the only metrics on the parameter space that monotonically decrease under an arbitrary completely positive and trace-preserving (CPTP) mapping $\calE$~\cite{Pe96}:
\e{
J^Q_{\bm\theta}(\{\h\rho_{\bm\theta}\})\ge J^Q_{\bm\theta}(\{\calE(\h\rho_{\bm\theta})\}).
}
From the monotonicity, the quantum Fisher information gives an upper bound of the classical Fisher information obtained by all the possible measurements, and the symmetric logarithmic derivative (SLD) Fisher information~\cite{He68}, which corresponds to $f(x)=\frac{1+x}{2}$, is known to be the least upper bound.
Therefore, the quantum Fisher information, especially SLD Fisher information, can be interpreted as the information on the parameter $\bm\theta$ that can be extracted from the quantum state.
We define the disturbance of the measurement as 
\e{
\Delta J^Q_{\bm \theta}:=J_{\bm \theta}^{Q}-\sum_{i}p_ {\bm\theta,i} J_{i,\bm \theta}'^{Q},\label{dist}
}
where $J_{\bm \theta}^{Q}$ and $J_{i,\bm \theta}'^{Q}$ are the quantum Fisher information of quantum statistical models $\{\h\rho_{\bm\theta}\}$ and $\{\h\rho_{\bm\theta,i}\}$, respectively.
For the sake of generality, we consider the disturbance using a general quantum Fisher information.

The information~\eqref{info} and the disturbance~\eqref{dist} satisfy the following inequality that shows a tradeoff relation between them:
\e{
J^C_{\bm\theta} \le  \Delta J^Q_{\bm\theta}.  \label{i-d}
}
This inequality means that if we perform a quantum measurement on an unknown state and extract information on the state, the state loses more intrinsic information.
Since the inequality~\eqref{i-d} is valid independently of the choice of the quantum Fisher information to define the disturbance, we obtain
\e{
J^C_{\bm\theta} \le \inf_Q \Delta J^Q_{\bm\theta},
}
where the infimum is taken over all kinds of the quantum Fisher information to define the disturbance.
We note that it is nontrivial which quantum Fisher information gives the minimum disturbance, though the minimum and the maximum of the quantum Fisher information are known to be the SLD Fisher information and the real right logarithmic derivative Fisher information (real RLD), which corresponds to $f(x)=\frac{2x}{x+1}$, respectively.

The proof of the inequality~\eqref{i-d} is based on the monotonicity and the separating property of the quantum Fisher information.
For a given measurement $\{\h K_{ij}\}$, we define a CPTP mapping $\calE^{\rm meas}$ as 
\e{
\calE^{\rm meas}(\h\rho)=\bigoplus_i \left( \sum_j \h K_{ij}\h\rho\h K_{ij}^\dagger \right).
}
Then the quantum Fisher information of the quantum statistical model $\{\calE(\h\rho_{\bm\theta})\}$ is equal to the sum of the classical Fisher information obtained by the measurement and the average quantum Fisher information of the post-measurement states as 
\e{
J^Q_{\bm\theta}(\{\calE^{\rm meas}(\h\rho_{\bm\theta})\})=J^C_{\bm\theta}(\{p_{\bm\theta}\})+\sum_{i}p_ {\bm\theta,i} J_{i,\bm \theta}'^{Q},  \label{separating}
}
which we call the separating property (see Appendix~\ref{proofA} for proof).
By applying the monotonicity under the CPTP mapping $\calE^{\rm meas}$, we obtain
\e{
J^Q_{\bm\theta}(\{\h\rho_{\bm\theta}\})&\ge J^Q_{\bm\theta}(\{\calE^{\rm meas}(\h\rho_{\bm\theta})\}) \nonumber \\
&=J^C_{\bm\theta}(\{p_{\bm\theta}\})+\sum_{i}p_ {\bm\theta,i} J_{i,\bm \theta}'^{Q},
}
which proves the inequality~\eqref{i-d}.

\section{condition for equality}
Measurements that achieves the equality of the inequality~\eqref{i-d} are efficient in a sense that they cause the minimum disturbance among those that extract a given amount of information.
When we adopt the right logarithmic derivative (RLD) Fisher information~\cite{YL73}, which corresponds to $f(x)=x$, to define the disturbance, a class of measurements called pure and reversible measurement achieves the equality of the inequality~\eqref{i-d} (see Appendix~\ref{proofB} for proof):
\e{
J^C_{\bm\theta}=\Delta J^{\rm RLD}_{\bm\theta} \label{RLDequality}
}
Here, a measurement is called pure if the number of measurement operators is one for each measurement outcome so that the measurement operators can be written as  $\{\h K_i\}$, and is called reversible if each $\h K_i$ has the inverse operator $\h K_i^{-1}$~\cite{UIN96}.

As an example, a measurement on a spin-1/2 system proposed by Royer~\cite{Ro94} is pure, whose measurement operators are 
\e{
\h K_1&=
\begin{pmatrix}
\cos(\theta/2-\sigma/4)&0\\
0&\cos(\theta/2+\sigma/4)
\end{pmatrix},\\
\h K_2&=
\begin{pmatrix}
\sin(\theta/2-\sigma/4)&0\\
0&\sin(\theta/2+\sigma/4)
\end{pmatrix}.
}
This measurement is reversible if $\theta/2\pm\sigma/4\ne n\pi/2$.

In fact, pure measurements are the least disturbing measurements in the following sense.
Suppose that two measurements $\{\h K'_{ij}\}$ and $\{\h K_i\}$ give the same positive operator-valued measure (POVM)
\e{
\sum_j \h K_{ij}'^\dagger\h K'_{ij}=\h K_i^\dagger \h K_i, \quad\forall i\in I,
}
and hence give the same probability distribution.
Then, the pure measurement causes less disturbance, and gives the same amount of information:
\e{
\Delta J^Q_{\bm\theta}(\{\h K_i\})&\le \Delta J^Q_{\bm\theta} (\{\h K'_{ij}\}),\\
J^C_{\bm\theta} (\{\h K_i\})&=J^C_{\bm\theta}(\{\h K'_{ij}\}).
}

\section{information-disturbance relation based on distinguishability}
In Ref.~\cite{BL05}, by using the classical and quantum relative entropies
\e{
S^C(p\|q)&=\sum_i p_i\log\left( \frac{p_i}{q_i} \right),\\
S^Q(\h\rho\|\h\sigma)&=\tr{\h\rho(\log\h\rho-\log\h\sigma)},\label{q.rel.}
}
a similar inequality to~\eqref{i-d} was derived:
\e{
S^C(p\|q)\le S^Q(\h\rho\|\h\sigma)-\sum_i p_i S^Q(\h\rho_i\|\h\sigma_i),\label{rel.ent}
}
where $p,q$ and $\h\rho_i,\h\sigma_i$ are the probability distributions and the post-measurement states of a quantum measurement performed on the quantum states $\h\rho, \h\sigma$, respectively.
Since the quantum relative entropy is a measure of the distinguishability of two quantum states~\cite{HP91,ON02}, Eq.~\eqref{rel.ent} can also be interpreted as a tradeoff relation between information and disturbance.
In particular, if we choose two similar states $\h\rho_{\bm\theta}$ and $\h\rho_{{\bm\theta}+{\rm d}{\bm\theta}}$ as the arguments of the relative entropy, Eq.~\eqref{rel.ent} reproduces the inequality~\eqref{i-d} with the disturbance defined by the Bogoliubov-Kubo-Mori (BKM) Fisher information, which corresponds to $f(x)=\frac{x-1}{\log x}$.

In the following, we discuss the extension of the inequality~\eqref{rel.ent} to general divergences.
Let $D^C(\cdot\|\cdot)$ be a divergence between two probability distributions, and $D^Q(\cdot\|\cdot)$ be its quantum extension, i.e.,
if two quantum states $\h\rho,\h\sigma$ commute and therefore are simultaneously diagonalizable as $\h\rho=\sum_ip_i\ket i\bra i, \h\sigma=\sum_iq_i\ket i\bra i$, we obtain
\e{
D^Q(\h\rho\|\h\sigma)=D^C(p\|q).
}
We note that quantum extensions of a divergence is not unique in general.

Let us consider a condition for divergences $D^C(\cdot\|\cdot),D^Q(\cdot\|\cdot)$ to satisfy the information-disturbance tradeoff relation
\e{
D^C(p\|q)\le D^Q(\h\rho\|\h\sigma)-\sum_i p_i D^Q(\h\rho_i\|\h\sigma_i).\label{divergence}
}
The essential properties for the proof of the inequality~\eqref{i-d} are the monotonicity and the separating property of the quantum Fisher information.
In the same way, we require these two properties for divergences:
\e{
D^Q(\h\rho\|\h\sigma)&\ge D^Q(\calE(\h\rho)\|\calE(\h\sigma)),\\
D^Q(\calE^{\rm meas}(\h\rho)\|\calE^{\rm meas}(\h\sigma))&=D^C(p\|q)+\sum_i p_iD^Q(\h\rho_i\|\h\sigma_i).
}
If we also require a continuity of $D^C(p\| q)$ with respect to $p,q$, then it satisfies Hobson's five conditions that axiomatically characterize the classical relative entropy~\cite{Ho69}.
Therefore, the divergence with the monotonicity, the separating property, and the continuity must be consistent with the relative entropy at least for classical probability distributions:
\e{
D^C(p\|q)=S^C(p\|q).
}
As is shown in~\cite{BL05}, the well-known quantum relative entropy satisfies the tradeoff relation~\eqref{divergence} because it has the monotonicity and the separating property.
Another quantum extension of relative entropy proposed by Belavkin and Staszewski~\cite{BS82},
\e{
S^{\rm BS}(\h\rho\|\h\sigma)=\tr{\h\rho\log(\h\rho^{1/2}\h\sigma^{-1}\h\rho^{1/2})}, \label{BS.rel.}
}
also satisfies the inequality~\eqref{divergence}.
Here, $S^{\rm BS}(\cdot\|\cdot)$ is known to be maximal among all the possible quantum extensions of the classical relative entropy~\cite{Ma13}.
By substituting $\h\rho_{\bm\theta}$ and $\h\rho_{{\bm\theta}+{\rm d}{\bm\theta}}$ to the inequality~\eqref{divergence}, we again obtain the inequality~\eqref{i-d} with the disturbance defined by the real RLD Fisher information.

\section{conclusion}
In this paper, we have formulated the tradeoff relation between information and disturbance in quantum measurement in view of estimating parameters that characterize an unknown quantum state.
The information is defined as the classical Fisher information of the probability distributions of measurement outcomes, and the disturbance is defined as the average loss of the quantum Fisher information due to the backaction of the measurement.
We have shown the tradeoff relation~\eqref{i-d} between them.
When we use the RLD Fisher information, the equality of the inequality~\eqref{i-d} is achieved by pure and reversible measurements.
In fact, pure measurements are the least disturbing among those that provide us with a given amount of information.

We have also discussed the necessary condition for divergences between two quantum states to satisfy a similar tradeoff relation~\eqref{divergence}.
It is necessary for divergences to coincide with the relative entropy at least for classical probability distributions.
In addition to the well-known relative entropy, the maximum relative entropy also satisfies the tradeoff relation~\eqref{divergence}, which reproduces the inequality~\eqref{i-d} for the disturbance defined by the real RLD Fisher information.
If there are quantum extensions of the relative entropy that give an arbitrary quantum Fisher information, another systematic derivation of the inequality~\eqref{i-d} should be possible.
\section*{acknowledgement}
This work was supported by
KAKENHI Grant No. 26287088 from the Japan Society for the Promotion of Science, 
a Grant-in-Aid for Scientific Research on Innovation Areas ``Topological Quantum Phenomena'' (KAKENHI Grant No. 22103005),
the Photon Frontier Network Program from MEXT of Japan,
and the Mitsubishi Foundation.
T. S. was supported by the Japan Society for the Promotion of Science through Program for Leading Graduate Schools (ALPS).
Y. K. acknowledges supports by Japan Society for the Promotion of Science (KAKENHI Grant No. 269905).

\appendix
\section{Proof of the separating property~\eqref{separating}}\label{proofA}
We introduce the logarithmic derivative operator of the quantum statistical model $\{\h\rho_{\bm\theta}\}$, defined as 
\e{
\h L_a:={\bm K}_{\h\rho_{\bm \theta}}^{-1} \left(\pd{1}{\h\rho_{\bm \theta}}{\theta_a}\right),
}
so that the quantum Fisher information is rewritten as
\e{
[J^Q_{\bm \theta}(\{\h\rho_{\bm\theta}\})]_{ab}:=\tr{\pd{1}{\h\rho_{\bm \theta}}{\theta_a}\h L_b}.
}

Let $\h L'_a$ and $\h L_{i,a}$ denote the logarithmic derivative of the quantum state models $\{ {\calE^{\rm meas}}(\h\rho_{\bm\theta})\}$ and $\{\h\rho_ {\bm\theta,i}\}$, respectively.
The relation between $\h L'_a$ and $\h L_{i,a}$ is calculated as
\e{
\h L'_a = \left(   \bigoplus_i \h L_{i,a} \right)
+\left(  \bigoplus_i \pd{1}{\log  p_ {\bm\theta,i}}{\theta_a} \h I_i     \right)  .
}
Therefore, the elements of the quantum Fisher information can be calculated as follows:
\e{
&\ \quad[J_{\bm \theta}^{\rm RLD}(\{\calE^{\rm meas}(\h\rho_{\bm\theta}\})]_{ab}\nonumber\\
&= \sum_i \left\{ p_ {\bm\theta,i}\tr{\pd{1}{\h\rho_{\bm\theta,i}}{\theta_a} \h L_{i,b}}  +  p_ {\bm\theta,i}   \pd{1}{\log  p_ {\bm\theta,i}}{\theta_b}  \tr {\pd{1}{\h\rho_{\bm\theta,i}}{\theta_a}}\right\}\nonumber\\
& \qquad +\sum_i \left\{ \pd{1}{p_ {\bm\theta,i}}{\theta_a} \tr{\h\rho_{\bm\theta,i}\h L_{i,b}}  +\pd{1}{p_ {\bm\theta,i}}{\theta_a}  \pd{1}{\log  p_ {\bm\theta,i}}{\theta_b} \tr{\h\rho_{\bm\theta,i}}  \right\}\nonumber\\
&= \sum_{i}p_ {\bm\theta,i} J_{i,\bm\theta}^{'Q}+0+0+J^C_{\bm\theta}.
}
Here, in obtaining the last equality, we use the fact that the econd and third terms vanish because
\e{
\tr {\pd{1}{\h\rho_{\bm \theta,i}}{\theta_a}}&= \pd{1}{}{\theta_a}\tr{\h\rho_{\bm \theta,i}}=0,\\
\tr{\h\rho_{\bm \theta,i} \h L_{i,b}}&= \tr{\h\rho_{\bm \theta,i}  {\bm K}^{-1}_{\h\rho_{\bm \theta,i}}\left(\pd{1}{\h\rho_{\bm \theta,i}}{\theta_b}\right)}\nonumber\\
&= \tr{{\bm K}^{\dagger-1}_{\h\rho_{\bm \theta,i}}(\h\rho_{\bm \theta,i})  \pd{1}{\h\rho_{\bm \theta,i}}{\theta_b}}=\tr{\h I_i  \pd{1}{\h\rho_{\bm \theta,i}}{\theta_b}}=0.
}

\section{Proof of Eq.~\eqref{RLDequality}} \label{proofB}
It is sufficient to prove 
\e{
J_{\bm \theta}^{\rm RLD}(\{\h\rho_{\bm\theta}\})=J_{\bm \theta}^{\rm RLD}(\{\calE(\h\rho_{\bm\theta}\})
}
because the disturbance can be rewritten as
\e{
\Delta J_{\bm \theta}^{\rm RLD}&= J_{\bm \theta}^{\rm RLD}(\{\h\rho_{\bm\theta}\})-J_{\bm \theta}^{\rm RLD}(\{\calE(\h\rho_{\bm\theta}\})+J^C_{\bm\theta}. \label{redef_disturbance}
}
Let $\h L_a$ and $\h L'_a$ denote the right logarithmic derivatives of the quantum state models $\{\h\rho_{\bm \theta}\}$ and $\{ {\calE^{\rm meas}}(\h\rho_{\bm\theta}) \}$.
Then, we obtain a simple relation between $\h L_a$ and $\h L'_a$ as follows:
\e{
\h L'_a&= \bigoplus_i (K_i\h\rho_{\bm\theta}\h K_i^\dagger)^{-1}\pd{1}{}{\theta_a}K_i\h\rho_{\bm\theta}\h K_i^\dagger \nonumber\\
&= \bigoplus_i (\h K_i^\dagger)^{-1}\h L_a\h K_i^\dagger.
}
Therefore,  we obtain
\e{
&\ \quad[J_{\bm \theta}^{\rm RLD}(\{\calE^{\rm meas}(\h\rho_{\bm\theta}\})]_{ab}\nonumber\\
&= \sum_i \tr{\h K_i \h\rho_{\bm\theta}\h K_i^\dagger (\h K_i^\dagger)^{-1}\h L_a  \h K_i^\dagger (\h K_i^\dagger)^{-1}  \h L_b\h K_i^\dagger }\nonumber\\
&= \tr{\sum_i \h K_i^\dagger\h K_i \h\rho_{\bm\theta}\h L_a\h L_b}\nonumber\\
&= \tr{\h\rho_{\bm\theta}\h L_a \h L_b}\nonumber\\
&= [J_{\bm \theta}^{\rm RLD}(\{\h\rho_{\bm\theta}\})]_{ab},
}
which completes the proof.

\bibliography{reference}

\end{document}